\documentclass[aps,prl,twocolumn]{revtex4}
\usepackage{graphicx}
\usepackage{amssymb}
\usepackage{hyperref}

\begin{document}

\title{Quantum phase transition in a two-dimensional system of dipoles}

\author{G.E.~Astrakharchik$^a$, J.~Boronat$^a$, I.L.~Kurbakov$^b$, Yu.E.~Lozovik$^b$}

\affiliation{$^a$ Departament de F\'{\i}sica i Enginyeria Nuclear, Campus Nord B4-B5, Universitat Polit\`ecnica de Catalunya, E-08034 Barcelona, Spain\\
$^b$ Institute of Spectroscopy, 142190 Troitsk, Moscow region, Russia}

\date{\today}

\begin{abstract}
The ground-state phase diagram of a two-dimensional Bose system with dipole-dipole
interactions is studied by means of quantum Monte Carlo technique. Our calculation
predicts a quantum phase transition from gas to solid phase when the density
increases. In the gas phase the condensate fraction is calculated as a function of
the density. Using Feynman approximation, the collective excitation branch is
studied and appearance of a roton minimum is observed. Results of the static
structure factor at both sides of the gas-solid phase are also presented. The
Lindeman ratio at the transition point comes to be $\gamma = 0.230(6)$. The
condensate fraction in the gas phase is estimated as a function of the density.
\end{abstract}

\maketitle

The chromium atom has exceptionally large permanent dipole moment and recent
realization of Bose-Einstein condensation of $^{52}$Cr\cite{CrBEC} has stimulated great
interest in properties of dipolar systems at low temperatures. It was
observed\cite{DipBEC} that dipolar forces lead to anisotropic deformation during
expansion of the condensate. In the experiments\cite{CrBEC,DipBEC}, the dipolar
forces were competing with short-range scattering. The latter, in principle, can be
removed by tuning the $s$-wave scattering length to zero by Feshbach
resonance\cite{Werner2005}. This would lead to an essentially pure system of
dipoles. On the other hand, low-dimensional systems can be realized by making the
confinement in one (or two) directions so tight, that no excitations of the levels
of the tight confinement are possible and the system is dynamically two- (or one-)
dimensional.

A major development has also been done in the present years towards the realization
of excitons at temperatures close to the Bose-Einstein condensation
temperature\cite{excitonsexper}. An exciton is much more stable if the electron is
spatially separated from the hole (spatially separated excitons). Such an exciton
can be modeled as a dipole. If the excitons are in two coupled quantum wells they
can be treated effectively as two-dimensional if the size of an exciton is small
compared to the mean distance between excitons.

One might expect to find a phase transition from gas phase to a crystal one at large
density. As the condensate fraction is small at the transition point, perturbative
theories, like Gross-Pitaevskii \cite{DipMF} or Bogoliubov
\cite{DipBogoliubov,Roton} approaches, will fail to describe accurately this
transition. One has to use {\it ab initio} numerical methods to address this quantum
many-body problem. Recently a trapped system of two-dimensional dipoles has been
studied by Path Integral Monte Carlo method\cite{Volkov} and mesoscopic analog of
crystallization has been found. Trapped dipoles with $s$-wave scattering were
investigated\cite{MonteCarloCr}.
So far, there have been no full quantum microscopic computations of the properties
of a homogeneous system of dipoles.

The Hamiltonian of a homogeneous system of $N$ bosonic dipoles has the form
\begin{eqnarray}
\hat H = -\frac{\hbar^2}{2M}\sum_{i=1}^N \Delta_i +
\frac{C_{dd}}{4\pi}\sum_{j<k}\frac{1}{|{\bf r_j - r_k}|^3},
\label{H}
\end{eqnarray}
where $M$ is the dipole mass and ${\bf r_i}, i=\overline{1,N}$ are the positions of
dipoles. The expression for the coupling constant $C_{dd}$ depends on the nature of
the dipole-dipole interaction. Two possible physical realizations of a
two-dimensional system of dipoles can be considered:

1) Cold atoms with permanent dipole moments $m$ aligned perpendicularly to the plane
of confinement by an external magnetic field. In this case $C_{dd} = \mu_0m^2$,
where $\mu_0$ is the permeability of free space. Alternatively, the electric dipole
moment can be induced by an electric field $E$, then the coupling constant is
$C_{dd} = E^2\alpha^2/\epsilon_0$, where $\alpha$ is the static polarizability and
$\epsilon_0$ the permittivity of free space.

2) Spatially indirect excitons. In two coupled quantum wells, one containing only
holes, and the other only electrons, holes and electrons might couple forming
indirect excitons. Another possible realization is a single quantum well where the
exciton dipole moment is induced by normal electric field. Spatial separation
between hole and electron suppress recombination and greatly increases the lifetime
of an exciton. If the separation between excitons are greater than the electron -
hole separation $D$, indirect excitons can be approximated as bosons with with
dipolar moment oriented perpendicularly to the plane. In this case $C_{dd} =
e^2D^2/\varepsilon$, where $e$ is an electron's charge and $\varepsilon$ is the
dielectric constant of the semiconductor. Study of 2D indirect exciton systems in
fact is a hot topic and this problem has been addressed both theoretically
\cite{excitonstheory} and experimentally\cite{excitonsexper}.

The Hamiltonian (\ref{H}) can be written in dimensionless form by expressing all
distances in units of characteristic length $r_0 = M C_{dd}/(4\pi\hbar^2)$ and
energies in units of ${\cal E}_0 = \hbar^2/Mr_0^2$. Properties of a homogeneous
system are governed by the dimensionless parameter $n r_0^2$ (dimensionless density)
with $n$ being density of the system.

In this paper we present a complete study of the phase diagram of two-dimensional
bosonic particles with dipole-dipole interactions at zero temperature. We resort to
the Diffusion Monte Carlo (DMC) method\cite{BORO1} in order to find the ground-state
energy and correlation functions of the many-body Hamiltonian (\ref{H}). Within DMC
method the Schr\"odinger equation is solved in imaginary time for the product of the
ground-state wave function and a trial (or guiding) wave function, which we chose of
Bijl-Jastrow form: $\Psi_T({\bf r_1},...,{\bf r_N}) = \prod_{i=1}^N f_1({\bf r_i})
\prod_{j<k}^N f_2({|\bf r_j-r_k}|)$. In the gas phase the density profile
is constant and $f_1(r) =$ const. When two particles closely approach, the influence
of other particles can be neglected and we expect that $f_2({\bf r})$ is well
approximated by the solution of the two-body scattering problem. We choose the short
distance part of the two-body correlation term as $f_2(r) = C_1 K_0(2/\sqrt{r})$,
where $K_0(r)$ is modified Bessel function of the second kind and $C_1$ is some
constant. At large distances, instead, contribution from other particles must not be
neglected and collective behavior (phonons) is expected. From hydrodynamic theory it
was shown \cite{Reatto} that in a two-dimensional system the phononic long-range
part of the wave function decays as $f_2(r) \propto \exp(-$const$/r)$. We use this
functional dependence in a symmetrized form $f_2(r) = C_2 \exp(-C_3/r)
\exp(-C_3/(L-r))$, which ensures that $f_2'(L/2)=0$. We match the short- and long-
range parts of $f_2(r)$ and demand continuity of the function and its first
derivative at a matching distance $R_{match}$. This, together with the condition
$f_2(L/2)=1$, fixes constants $C_1,C_2,C_3$. In the solid phase we use the same
$f_2(r)$ and add the one-body Gaussian localization term $f_1(r_i) = \exp[-\alpha
({\bf r_i - r_i^{cr}})^2], i=\overline{1,N}$, where $\alpha$ is the localization
strength and $r_i^{cr}$ position of the lattice site. Variational parameters
$R_{match}$ and $\alpha$ are chosen by minimizing the variational energy.

We model the homogeneous system at density $n$ by considering $N$ particles in a
simulation box with periodic boundary conditions. The size of the box $L_x \times
L_y$ is chosen in such a way that $n = N /(L_xL_y)$. In the gas phase a square box
$L_x = L_y$ is considered, while in the solid phase we ensure that each of the box
sides is a multiple of an elementary cell size in a triangular lattice.

\begin{figure}
\begin{center}
\includegraphics[angle=0,width=0.95\columnwidth]{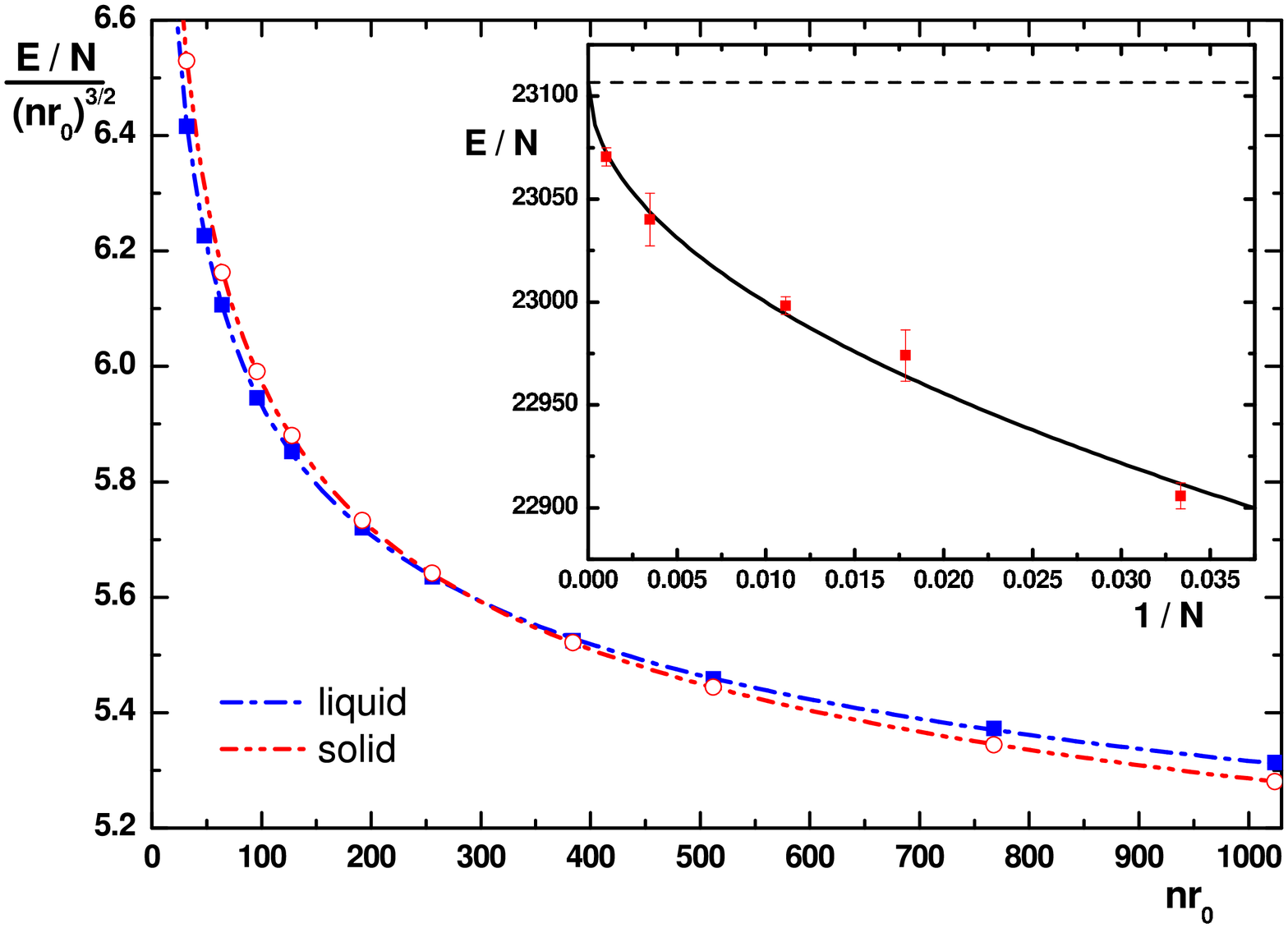}
\caption{(color online) Energy per particle of the dipole system as a function of the
dimensionless density $n r_0^2$: (blue) solid squares -- gas phase, (red) open
circles -- solid phase. Energy is measured in units $[\hbar^2/(Mr_0^2)]~/~(nr_0^2)^{3/2}$.
Inset: an example of the finite-size dependence for the energy in the solid phase at
dimensionless density $n r_0^2 = 256$, symbols - DMC energy (with added tail energy,
Eq.~(\ref{Etail})), solid line -- fit $E_{th} + C/\sqrt{N}$, dashed line -- extrapolation
of the energy to infinite system size $E_{th}$. Energy in the inset is measured in
units of $\hbar^2/(M r_0^2)$.}
\label{Fig:E}
\end{center}
\end{figure}

The advantage of DMC method is that it gives essentially exact energy within some
statistical uncertainties. In a system of dipoles special attention should be paid
to an appropriate extrapolation to the infinite system. The reason for that is that
while $1/r^3$ dipole-dipole interaction is not yet a long-range one in a
two-dimensional system \cite{Ruffo}, the long-range decay is already quite slow.
The finite-size effects in energy can be significantly reduced by adding the {\it
tail} energy
\begin{eqnarray}
\frac{E_{tail}(n,L)}{N} = \frac{1}{2} \int_{L/2}^\infty V(r)g_2(r) 2\pi r dr
\label{Etail}
\end{eqnarray}
to the output of the DMC calculation with $L=\min\{L_x,L_y\}$. Here $V(r)$ is the
interaction potential and $g_2(r)$ is the pair distribution function. An approximate
value of the integral (\ref{Etail}): $E_{tail}/N = C_{dd} n^{3/2}/(2\sqrt{N})$ is
obtained by substituting $g_2(r)$ by its average value in the bulk $g_2(r)=n$. This
greatly suppresses finite-size dependence (for the example shown in the inset of
Fig.~\ref{Fig:E} by a factor of 25), while the residual dependence is eliminated by
a fitting procedure. We note that the dependence on the number of particles in this
case scales as $1/\sqrt{N}$, contrary to the law $1/N$, usual for fast decaying
potentials.
We
do the extrapolation of the energy $E(N) = E_{DMC}+E_{tail}$ to the thermodynamic
limit value $E_{th}$ using as fitting formula: $E(N) = E_{th} + C/N^{1/2}$, where
$C$ is a fitting constant. In the inset of Fig.~1 we show an example of the finite
size study for the energy. In it we consider a density $nr_0^2=256$ and solid phase,
where we expect to find largest finite-size dependence, due to the oscillatory
behavior of the pair distribution function. We find that our fitting function
describes well the finite-size dependence and we use it for the extrapolation to the
thermodynamic limit. The same procedure is repeated for densities
$nr_0^2=32;48;64;96;128;196;256;384;512;768;1024$ and for the gas and solid
phases.

In Fig.~\ref{Fig:E} we show the results of the ground-state energy for gas and solid
phases. We find that at small density the gas phase is energetically favorable and
the solid phase is metastable. At larger densities the system crystallizes and a
triangular lattice is formed. We fit our data points with a dependence $E/N = a_1
(nr_0^2)^{3/2}+a_2 (nr_0^2)^{5/4}+a_3 (nr_0^2)^{1/2}$, where powers $3/2$ and $5/4$
describe proper asymptotic behavior of potential and kinetic energies respectively
and power $1/2$ is added for a better accuracy of the fit. The best parameters of
the fit are $a_1=4.536(8); a_2=4.38(4); a_3=1.2(3)$ for the gas phase and $a_1 =
4.43(1); a_2 = 4.80(3); a_3 = 2.5(2)$ for the solid. The transition is estimated to
happen at the critical density $nr_0^2=290(30)$. We note that in a one-dimensional
system the role of interactions is enhanced and the transition happens at much
smaller densities $nr_0 \approx 0.4$ \cite{1Ddip}. Maxwell double tangent
construction shows that the region of phase coexistence is very small and freezing
and melting points are indistinguishable within the error bars of our calculation.
At large densities quantum fluctuations get suppressed and the energy is dominated
by the potential energy of particle-particle interactions. The energy eventually
approaches the limit of a classical crystal. The triangular lattice in
this limit has potential energy $E_{triang}/N = 4.446 (nr_0)^{3/2}$. 

\begin{figure}
\begin{center}
\includegraphics[angle=0,width=0.95\columnwidth]{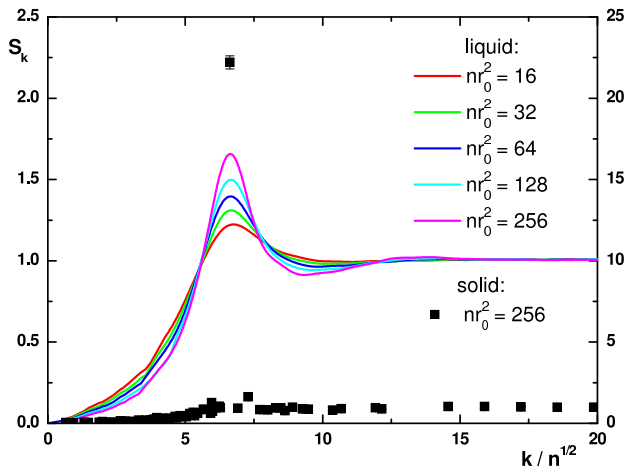}
\caption{(color online) Static structure factor: left axis -- gas phase at
densities $n r_0^2 = 16, 32, 64, 128, 256$ (the higher is the peak, the larger is the
density); right axis -- solid phase at density $n r_0^2 = 256$}
\label{Fig:Sk}
\end{center}
\end{figure}

Knowledge of the equation of state $E(n)/N$ of a homogeneous system permits the
calculation of the chemical potential $\mu(n) = \partial E/\partial N$ and the speed
of sound $c$: $Mc^2 = n \partial \mu/\partial n$. This information is extremely
useful for the description of trapped systems with large number of particles. As is
well known, the local density approximation can be used for predictions of the
density profile, release energy, frequencies of collective modes, {\it etc.} which
can be accessed in experiments \cite{LDA}.

We estimate the thermodynamic Lindemann ratio $\gamma = \sqrt{\langle({\bf
r-r^c})^2\rangle}/a_L$, ($a_L$ is the lattice length), at the transition density as
$\gamma = 0.230(6)$. Comparing to other two-dimensional systems we find that this
value is smaller than the one of a hard-disks $\gamma = 0.279$
, while is similar to more long-ranged potentials, like $\gamma = 0.235(15)$ for
Yukawa bosons
 and $\gamma = 0.24(1)$ for Coulomb bosons
\cite{Lind2D}. In a three-dimensional system the value of $\gamma$ at transition is
generally larger; for example $\gamma = 0.28$ for the Yukawa potential
\cite{Ceperley78}.

The breaking of translational symmetry takes place in the crystal. Periodic
modulation appears in the density profile $n(r)$ and the pair correlation function
$g_2(r-r') = \langle n(r)n(r')\rangle$. We have studied the static structure factor
$S_k$ which is related to $g_2(r)$ by Fourier transformation $S_k = 1+
2\pi\int_0^\infty (g_2(r)-n) J_0(kr)r\;dr$. We use a technique of pure estimators
which essentially removes any bias from a particular choice of the guiding wave
function\cite{pure}. The static structure factor increases from zero at $k=0$ to
unity for large momentum. In the gas phase $S_k$ increases smoothly and has a peak
around inverse interparticle distance $n^{1/2}$ (see Fig.~\ref{Fig:Sk}). As the
dimensionless density $nr_0^2$ is increased the correlations get stronger and the
height of the peak is increased. In the solid phase static structure factor is
discontinuous and has a $\delta$-function peak at inverse lattice site length
$a_L^{-1} = (\sqrt{3}n/2)^{1/2} = 0.93... n^{1/2}$.

\begin{figure}
\begin{center}
\includegraphics[angle=0,width=0.95\columnwidth]{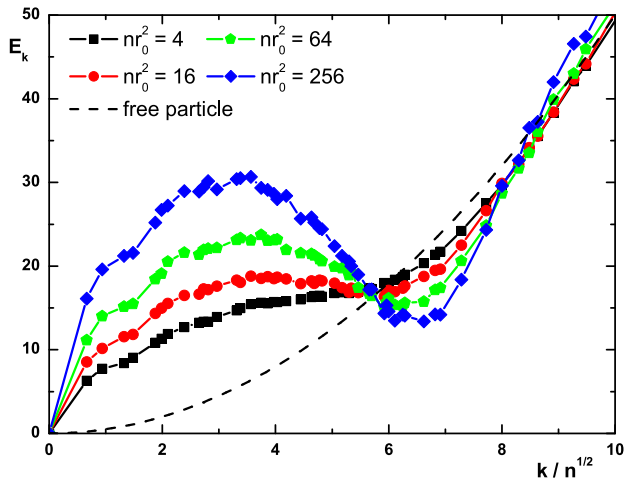}
\caption{Upper bound for the excitation spectrum in the
gas phase. Solid lines -- Feynman formula, Eq.~(\ref{Feynman}), applied to the
DMC data for the static structure factor, dashed line -- free particle limit. Energy
is measured in units of $\hbar^2/(M r_0^2)$.}
\label{Fig:spectrum}
\end{center}
\end{figure}

We can get some insight on the excitation spectrum $E_k$ by relating it to the
static structure factor through the Feynman relation
\begin{eqnarray}
E_k = \frac{\hbar^2k^2}{2M S_k}.
\label{Feynman}
\end{eqnarray}
This relation is expected to be exact for small momenta ({\it i.e.} in the phononic
regime), while at larger momenta it gives an upper bound to the excitation spectrum
\cite{SkUpperBound}. Predictions for $E_k$ obtained from DMC data are presented in
Fig.~\ref{Fig:spectrum}. We find that deviations from low-momenta phononic linear
behavior appear very soon. The excitation spectrum is monotonous for the smallest
considered density. As the density increases, the roton minimum at finite value of
the momentum is observed (see, also Refs.~\cite{Roton}). The roton gap $\Delta$
decreases as the density is increased. We expect that our approach gives the correct
position of the roton minimum, while the real value of the roton gap $\Delta$ is
overestimated, as it happens in DMC calculations for gas helium \cite{BoronatSk}.

\begin{figure}
\begin{center}
\includegraphics[width=0.95\columnwidth]{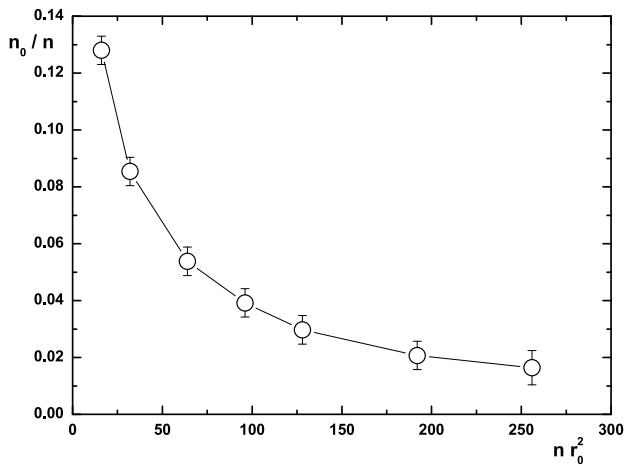}
\caption{Condensate fraction as a function of the dimensionless density $nr_0^2$.}
\label{Fig:CF}
\end{center}
\end{figure}

While Bose-Einstein condensation (BEC) is expected to happen in a dilute systems,
strong interactions between particles destroys coherence. In order to study the
process of decoherence quantitatively we measure condensate density $n_0$. In a
homogeneous system it can be found from the long-range asymptotic $|r-r'|\to\infty$
of one-body density matrix (OBDM)
$g_1(r-r')=\langle\hat\Psi^\dagger(r)\hat\Psi(r')\rangle$, where $\hat\Psi(r)$ is
the field operator. As $g_1(r)$ is a non-local quantity, the DMC output gives for it
a {\it mixed} estimator, which is biased by the choice of the guiding function. This
bias can be significantly diminished by the extrapolation procedure\cite{pure}. We
have measured asymptotics of OBDM for different sizes of the system and made
extrapolation to the thermodynamic limit. In Fig.~\ref{Fig:CF} we plot the
condensate fraction as a function of the dimensionless density in the gas phase. We
find that the condensate depletion is large in the range of considered densities.
This prohibits use of perturbative Bogoliubov approach for prediction of the
condensate depletion and justifies the necessity of a numerical approach to the
problem. The condensate fraction is rather small at the transition density.

In conclusions, we have investigated the zero-temperature quantum phase diagram of a
two-dimensional dipole system and estimated the critical density of the gas to solid
phase transition. We studied structural properties by observation of the pair
distribution function, static structure factor, Lindemann ratio. Coherence
properties are investigated by looking at the one-body density matrix. The
condensate fraction decreases rapidly with the density and is estimated to be very
small ($2\%$) at the transition point. Our predictions can be checked in future
experiments with spatially separated indirect excitons and cold dipolar Bose gases
in reduced geometry.

After completion of this work, we became aware of Ref. \cite{Pupillo}
in which systems of N=36,90 dipoles are studied by finite-temperature
Path Integral MC method with a result for $n_c$ compatible with our
prediction.

Some of the authors (G.E.A. and J.B.) acknowledge support by (Spain) Grant No.
FIS2005-04181 and Generalitat de Catalunya Grant No. 2005SGR-00779. The work was
partially supported by RFBR.

\end{document}